\documentclass{article}
\usepackage{spconf,amsmath,graphicx}
\usepackage{cite}
\usepackage{array}
\usepackage{multirow}
\usepackage{balance}
\usepackage{enumerate}

\newcolumntype{x}[1]{%
>{\centering\hspace{0pt}}p{#1}}%

\title{Atmospheric turbulence removal using convolutional neural network}
\name{Jing Gao, N. Anantrasirichai and David Bull}
\address{Visual Information Laboratory, University of Bristol, UK}

\begin{document}
%
\maketitle
\begin{abstract}
This paper describes a novel deep learning-based method for mitigating the effects of atmospheric distortion. We have built an end-to-end supervised convolutional neural network (CNN) to reconstruct turbulence-corrupted video sequence. Our framework has been developed on the residual learning concept, where the spatio-temporal distortions are learnt and predicted.  Our experiments demonstrate that the proposed method can deblur, remove ripple effect and enhance contrast of the video sequences simultaneously. Our model was trained and tested with both simulated and real distortions. Experimental results of the real distortions show that our method outperforms the existing ones by up to 3.8\% in term of the quality of restored images, and it achieves faster speed than the state-of-the-art methods by up to 23 times with GPU implementation.
\end{abstract}


\section{Introduction}
\label{sec:intro}

\begin{figure}[t!]
	\centering
  		\includegraphics[width=\columnwidth]{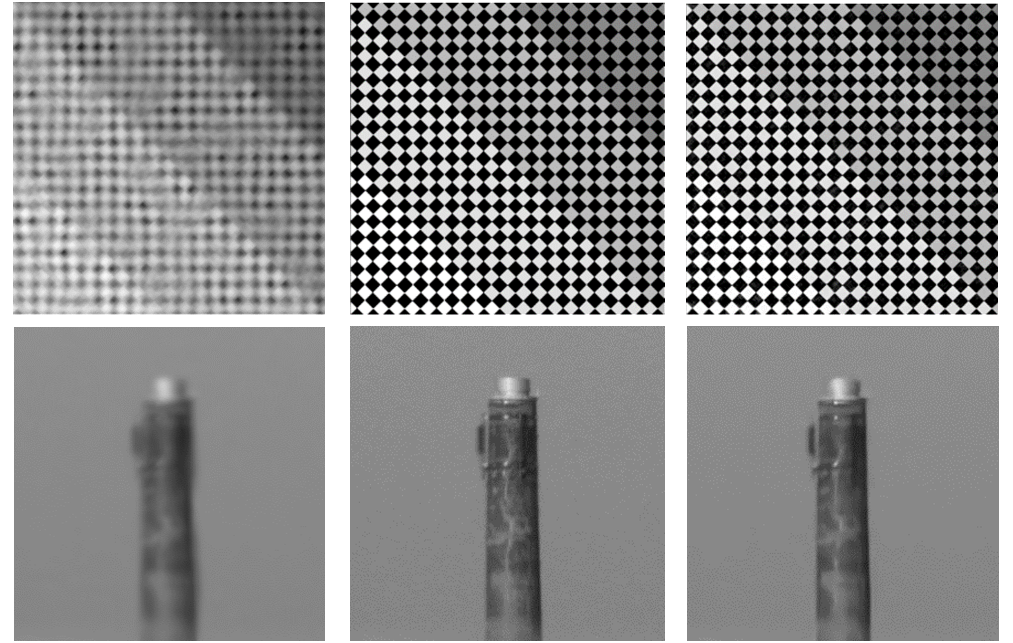}
	\caption{Restoration of atmospheric turbulence distortion from (top) simulated turbulence on a chessboard and (bottom) real scene \cite{5540158}. (Left to right) Distort image, ground truth and our restored image.}
	\label{fig:exampleresults}
\end{figure}

Distortions due to atmospheric conditions can degrade the visual quality of image during acquisition from cameras. This phenomenon can be found over hot roads, deserts and the object around flames, where the atmospheric turbulences appear as spatio-temporal ripples in video sequences (as shown in Fig.\ref{fig:exampleresults} left column). This phenomenon also reduces contrast, sharpness and the capacity to see objects at a long distance, such as those in surveillance applications. It often combines with other detrimental effects, like fog, which make the acquired imagery even more difficult to interpret. 

Previous work on atmospheric turbulence removal, e.g. complex steerable pyramid (CSP) \cite{zhang2018removing} and complex wavelet transform (CW) \cite{Anantrasirichai:mitigating:2012,Anantrasirichai:Atmospheric:2013}, is normally based on image processing methods when referring to software-based techniques.
These methods usually have three major drawbacks. Firstly, they generally involve a complex optimisation process which needs large buffers, computational processors \cite{Zhang:DnCNN:2017,Anantrasirichai:Atmospheric:2013} and long computational time \cite{anantrasirichai2018atmospheric, lucas2018using}. Secondly, they usually deal with geometric distortion and blurry degradation separately as they were considered as irrelevant problems, which increased the total complexity. Thirdly, the complex models could be non-convex and some parameters need to be chosen manually \cite{Anantrasirichai:Atmospheric:2013}. Based on these issues, we hence consider exploring a deep learning method to reconstruct the distorted images. Here, a convolutional neural network (CNN) is employed to characterise atmospheric turbulence distortions in video sequences, and then the sporadic motions and rapid distortions will be mitigated with locally trained filters in each layer of the network. Moreover, CNN-based methods have proved that, through implementation using graphics accelerator devices, they can achieve real-time processing and require reduced memory \cite{8100173}.


Image restoration can be addressed as an inverse problem and deep learning techniques have been employed to solved it. The learning-based image restoration for a specific distortion  includes deblurring \cite{nah2017deep}, denoising \cite{chen2018image} and super-resolution \cite{dong2015image}. More details of deep learning for inverse problems can be found in \cite{lucas2018using}.
The atmospheric turbulence scenario, however, contain blur effects combined with small-scale intensity fluctuations due to scintillation, which means this problem comprises several distortions. As such, existing deep learning methods are possibly not suitable for removing atmospheric turbulence.

This paper presents a new method to deal with atmospheric turbulence effects on images using a deep neural network. To the best of our knowledge, this work is the first to apply deep learning to the task of alleviating turbulence effect on images. We propose to restore atmospheric turbulence-corrupted images using the modified residual learning of deep CNN (DnCNN) \cite{Zhang:DnCNN:2017}. Here, the DnCNN, originally introduced for image denoising, is adapted to remove noise and spatio-temporally invariant blur simultaneously. It employs residual learning with batch normalization to achieve better image quality and faster training speed. We adapted the size of filters from 3$\times$3 to $n \times n$, where $n>3$, to obtain larger receptive field and more adjustable parameters for this more complicated task. We also investigate the performance of our model when using a single distorted image and an average of the multiple distorted images as the input of the network.

\begin{figure}[t!]
	\centering
  		\includegraphics[width=\columnwidth]{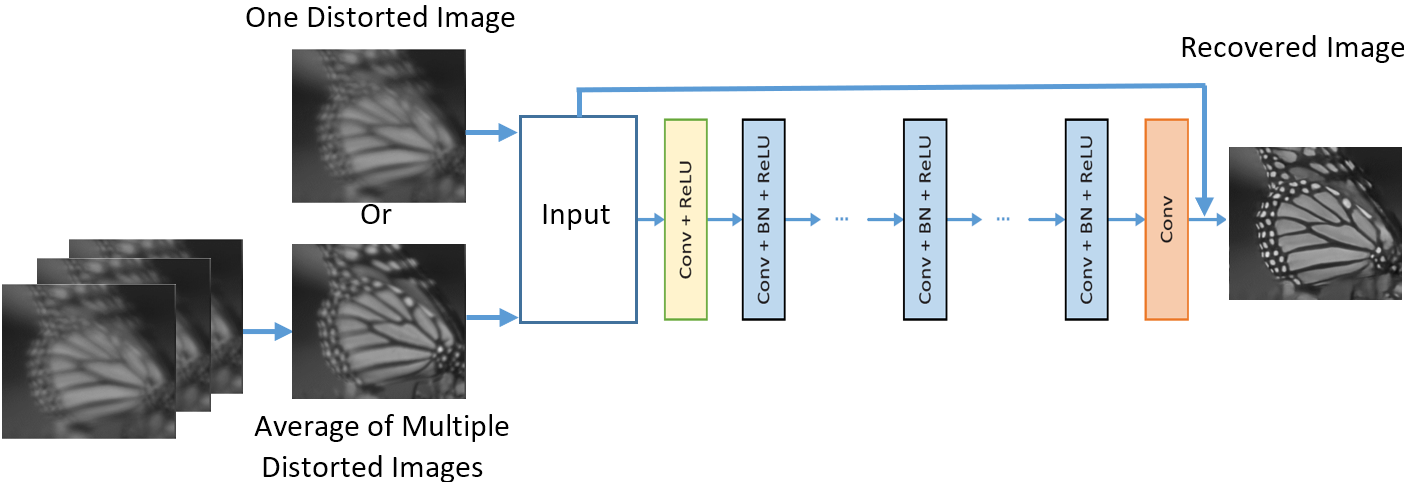}
	\caption{The proposed architecture adapted from DnCNN \cite{Zhang:DnCNN:2017}, showing two options of input, a distorted image or the average of multiple distorted frames.}
	\label{fig:arch}
\end{figure} 

\section{Deep learning for Atmospheric turbulence removal}
\label{sec:proposedscheme}

Atmospheric turbulence mitigation can be seen as an inverse problem, solving degradation model \cite{Anantrasirichai:Atmospheric:2013}, $y=hx+b$,
where $y$ represents the distorted observation of an ideal image  $x$ (ground truth). $h$ represents geometric distortion and blurry degradation, whilst $b$ is noise from sensors or transmission process. Our target is to recover $x$ from $y$. A direct mapping between ground truth and observed distorted image is $F(y; \theta)=x$, where $F$ is the mapping function and $\theta$ represents a set of parameters of $F$. The key is to find $F$ and its $\theta$.
CNN regards image pairs ($x$, $y$) as the inputs for training; each $x_i$ and each $y_i$ are a pair. Then CNN updates $\theta$ that include weights of filters and parameters of batch normalization through backpropagation by reducing the difference between $y$ and $x$.

Recently a residual learning has proved that it can accelerate the training process by learning a residual mode\cite{Zhang:DnCNN:2017}. In this paper, we use residual learning formulation to train a residual mapping $R(y; \theta)$ between the deformation map (residual image) and the distorted input, i.e. $x=y-R(y; \theta)$. Eq.\ref{ref:lossfn} shows the loss function to learn the trainable parameters $\theta$ for this mapping. The loss function is the sum square value of each pixel between ground truth images and recovered images.
{\setlength\abovedisplayskip{1pt}
\setlength\belowdisplayskip{1pt}
\begin{equation}
	 L( \theta)=\frac{1}{2m} \sum _1^m \left(|R(y_i; \theta) -(y_i-x_i)|^2 \right)
	 \label{ref:lossfn}
\end{equation}}where $L$ is a loss function calculated by mean square error, and $m$ is the number of image pairs.

\subsection{Network architecture}
The modified architecture, shown in Fig. \ref{fig:arch}, contains three kinds of layers.
\begin{enumerate}[i)]
 \item \textit{Convolution + ReLU}: In the first layer, 64 filters (size $n \times n$) are used for generating 64 feature maps through convolution. Then ReLU is applied for non-linearity. It is used in the first layer (yellow block in Fig. \ref{fig:arch}). The size of the filter ($n \times n$) is defined to be large enough to cover the spatial movement of the ripples. However, this value is unknown in real scenes. In this paper we manually measured it and define $n$ associated to the receptive field, explained in the next subsection. For future work, an optical flow technique could be employed to estimate the maximum pixel motion \cite{Chen:registation:2011}.
 \item \textit{Convolution + Batch normalization + ReLU}: From the second to the sixteenth layer, 64 filters with the size of $n \times n \times$64 are used. There is batch normalization between convolution and ReLU. As batch normalization has little influence on the front layers, we only add batch normalization in the hidden layers (blue blocks in Fig. \ref{fig:arch}).
 \item \textit{Convolution}: In the last layer (orange block in Fig. \ref{fig:arch}), one filter with the size $n \times n \times$64 is implemented to output a deformation map. 
 \end{enumerate}
 Finally, the recovered image is created by subtracting the deformation map to the distorted image.

\subsection{Network depth}
Deep depth and large convolutional filters generally give better feature extraction performance. However, they require more parameters and thus may bring in redundancy. In this paper, we use 17 layers to balance between efficiency and performance, similar to the architecture in \cite{Zhang:DnCNN:2017}. We however leave the filter size open to $n \times n$ so that it can be freely defined to make the reception field  cover the range of pixel motion according to the turbulence effects in different scenes. For severe turbulence, the pixel motion requires a larger effective receptive field.  In our dataset, we observe that the maximum pixel displacement is less than 30 pixels in a single direction (total moving range is 60 pixels). The filter size of 5$\times$5 ($n$=5) with a depth of 17 results the receptive field to 69$\times$69 pixels, which cover the pixel displacement of the turbulence. We have tried the original filter size (3$\times$3) of DnCNN, and it gave worse results for removing atmospheric turbulence.
A large size of training patches is required as it helps to make maximum use of context information, leading to the better recovering performance. Here, we set the patch size of 80$\times$80 pixels, which is slightly larger than the receptive field. 

\subsection{Input of the network}
The atmospheric turbulence can be viewed as being quasi-periodic; therefore, averaging a number of frames yields a geometric improvement in the image, but it remains blurred by an unknown point spread function (PSF) of the same size as the pixel motions due to the turbulence \cite{Mao:nonrigid:2012}. In this paper, we investigate the performance of our model when using a single distorted image and using the average of group of the distorted images as the input of the network. This average approach can be simply applied to the distorted sequence with moving objects by using sliding temporal window approach.

\section{Results and discussion}
\label{sec:results}

We tested our proposed method with both simulated and real datasets. 
The model was trained by with Adam optimizer with batch size of 128 and learning rate from 1e-3 to 1e-5. 
The batch normalisation is implemented to further stable the parameters. Same padding is implemented in all convolution layers. We set the convolution stride equals to 1. The training was done over 1000 epochs.  
The CPU in this paper is with Intel Core i5-6200U @2.30GHz and the GPU is Nvidia Tesla P100-PCIE-16GB.


\subsection{Experiment on synthetic datasets}

\begin{figure}[t!]
	\centering
  		\includegraphics[width=\columnwidth]{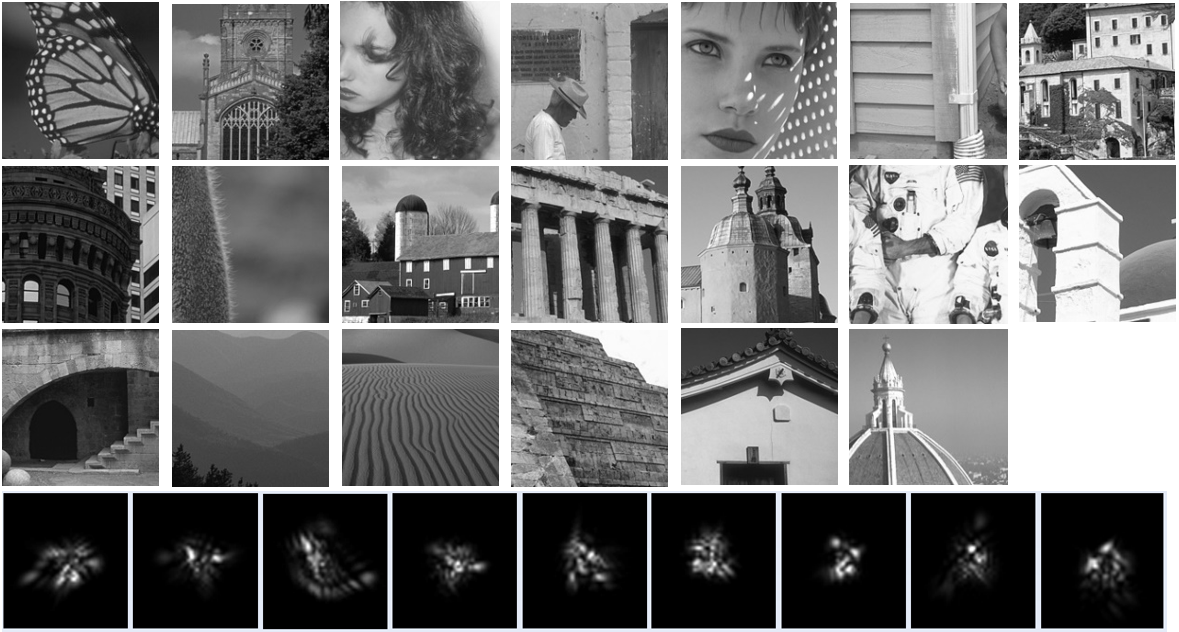}
	\caption{Row 1-3 show clean images of simulated datasets gathered from \cite{Zhang:DnCNN:2017}. Row 4 shows nine different atmospheric PSFs used for generated the synthetic datasets  \cite{5540158}.}
	\label{fig:dataset}
	\vspace{3.5mm}

	\centering
  		\includegraphics[width=\columnwidth]{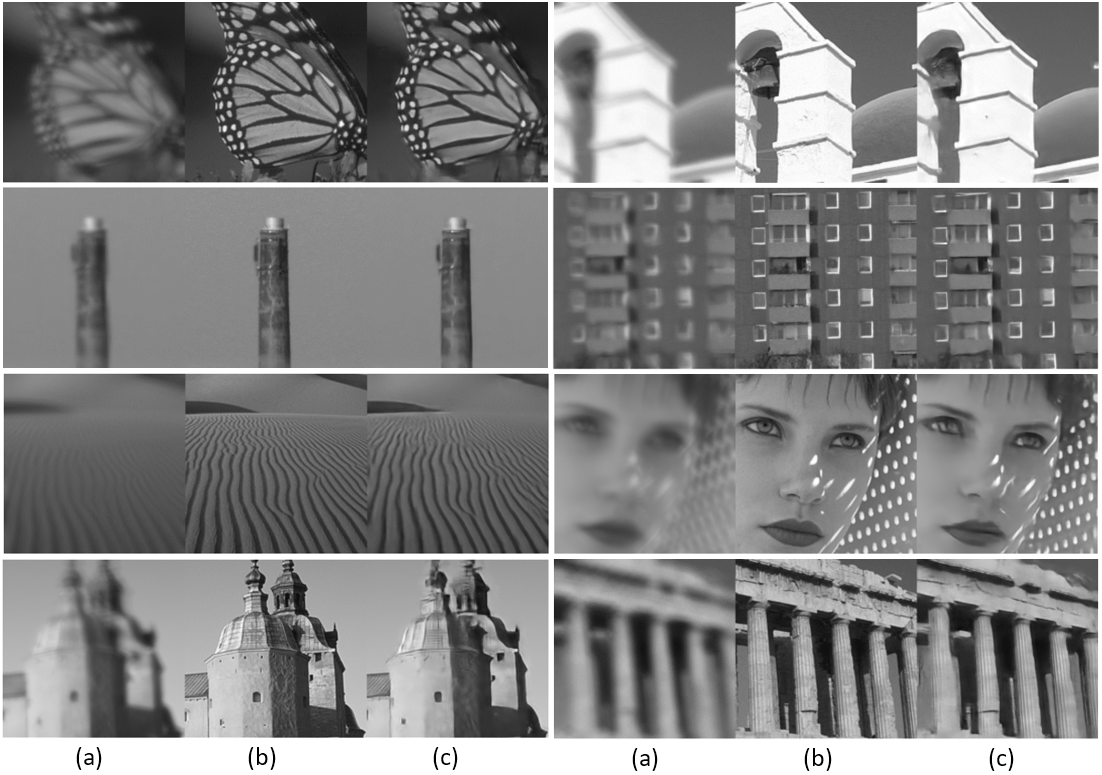}
	\caption{Reconstruction of simulated dataset. (a) distorted image, (b) ground truth, (c) recovered image.}
	\label{fig:simul}
\end{figure}

Datasets with ground truth is essential for supervised learning; unfortunately, the availability of the datasets for atmospheric turbulence problem is very limited. Therefore, we generated the synthetic atmospheric turbulence distortion. This was done with 9 PSFs provided in \cite{5540158}. The spatially variant blur is created by applying the randomly selected PSF to the different parts of the image, and for each image the PSFs are resized randomly so that the strength of ripple effects and degree of blur vary between frames. Then a Gaussian noise is added. 
We employed 20 clean images with a size of 180$\times$180 pixels \cite{Zhang:DnCNN:2017} shown in Fig. \ref{fig:dataset} 
and 2 clean images: `building' and `chimney' ground truth provided in \cite{5540158}. 



We generated 2 separate synthetic datasets (Fig. \ref{fig:simul}(a)) for training and testing the model. These images share the same ground truth but with different degrees of distortion. The result for recovering a simulated image with a single distorted image is presented in Fig. \ref{fig:simul} (c). We can see that most blurs and distortions have been removed for all images. Table \ref{tab:performance_simul} shows the PSNR and SSIM of 6 scenes in Fig. \ref{fig:simul}. All PSNR of the restored results are over 24dB and all SSIM are over 0.76. We can see that the proposed method can improve the image quality by up to 30\% from the original quality.

\begin{table}[t!]
	\centering
	\caption{Objective assessment of simulated distorted dataset ($d$) and the restored results ($r$)}
	\small
	\scalebox{0.91}{
		\begin{tabular}{ccccccc}
		\hline
			\multirow{1}{*}{Image} &\multirow{1}{*}{Butterfly} & \multirow{1}{*}{Arch} & \multirow{ 1}{*}{Chimney} &\multirow{1}{*}{Building} & \multirow{1}{*}{Desert} & \multirow{1}{*}{Face} \\
			\hline
			\multirow{ 1}{*}{PSNR$_{d}$}	& 19.27& 19.82 & 32.77& 23.49 &24.81&21.48 
 \\
			\multirow{ 1}{*}{SSIM$_{d}$}	& 0.61 & 0.73 & 0.95 &0.78 &0.65& 0.73 \\
			\hline 
			\multirow{ 1}{*}{PSNR$_{r}$}	 & 24.28 & 	26.88	& 39.49	 &28.45  & 25.64&28.39\\
            \multirow{ 1}{*}{SSIM$_{r}$}	& 0.88 & 0.93&	0.98&0.92& 0.76&0.86\\
			\hline 
		\end{tabular}}
	\label{tab:performance_simul}
\end{table}

\subsection{Experiment on static real datasets}



We employed the test datasets, `building' and `chimney', captured from real word  \cite{5540158}. Each scene contains 100 frames, and the first 80 were used for training. Test results of the state-of-the-art methods were gathered from their published papers directly. Our method mitigates the distortion in frame-by-frame basis, i.e. one restored image is outputted for one distorted image. This is different from other existing methods, where multiple inputs are required to generate one restored image. Therefore our objective results are shown as an average of PSNR and SSIM of 20 testing images ($\bar{y}_{20}$), when only single distorted input is use.
However, the visual result of the `chimney' can not always get result as good as that in Fig. 1, because the real distortion contain more complicated distortion that we may not learn well with the limited dataset. To seek for a stable result, we investigated the performance of our model for multiple inputs by simply averaging these distorted images as the input  (as shown in Fig. \ref{fig:arch}), and the result is one restored frame ($\bar{x}$).

\begin{figure}[t!]
	\centering
  		\includegraphics[width=\columnwidth]{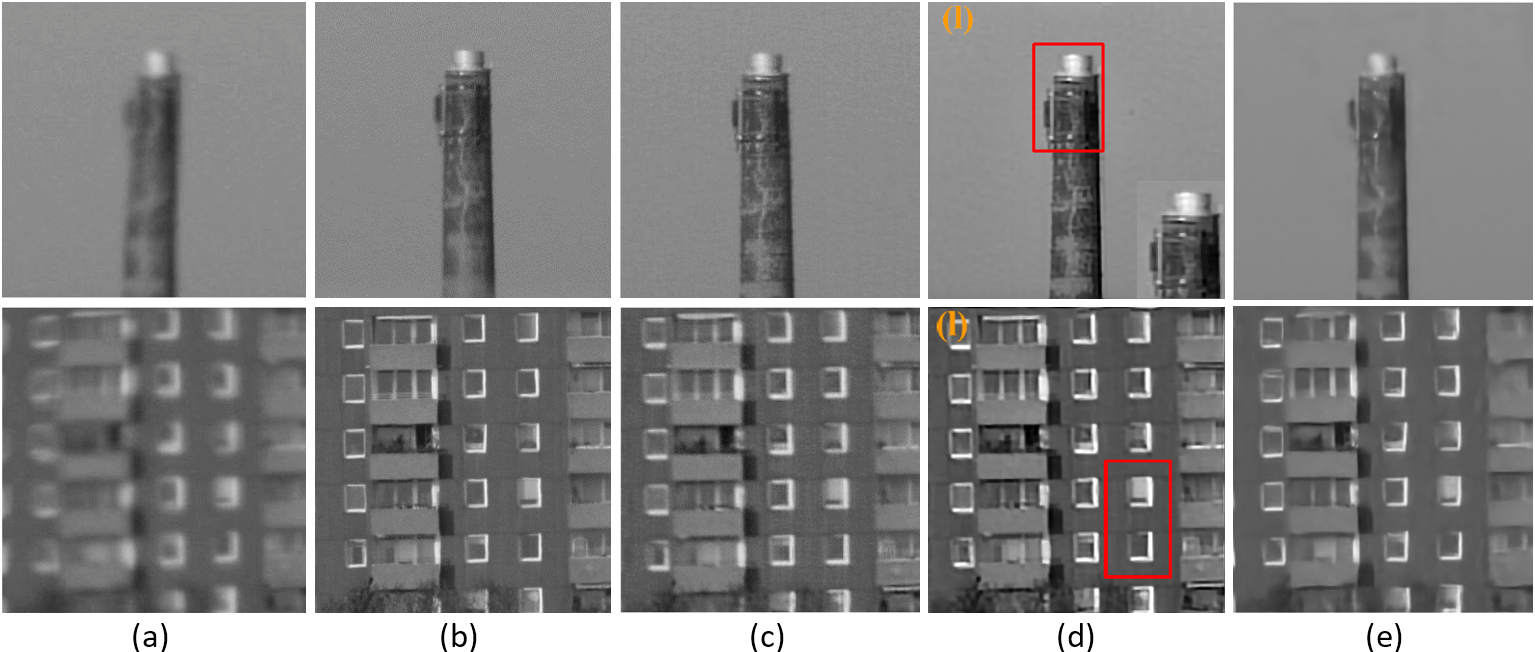}
	\caption{Comparison of results on the real dataset. (a) distorted image, (b) ground truth, (c) CW, (d) CSP, (e) proposed method. CW and CSP are reconstructed from selected frames, whilst the input of our method is the average of 20-30 images.}
	\label{fig:re}
\end{figure}

The subjective results are shown in Fig. \ref{fig:re}, where (a) are the distorted images, (b) are ground truth, (c) and (d) are the reconstruction of CW \cite{Anantrasirichai:Atmospheric:2013} and CSP \cite{zhang2018removing} separately, and (e) are the proposed method. CW and CSP are reconstructed from a number of good frames or patches selected from 100 frames, whilst our method employed only 30 and 20 continuous input frames and averaged them for `chimney' and `building' separately. CSP and CW can derive sharp results, but the visual quality may be different from the ground truth. 

The corresponding objective results of Fig. \ref{fig:re} are shown in Table \ref{tab:performance}. The best PSNR and SSIM are highlighted in bold. We can see that the proposed method with the average input ($\bar{x}$) yields the highest PSNR. Specifically, it outperforms the competing methods by 1.66dB (CW) and 2.83 dB (CSP). But the SSIM value for building is the highest for CW.  

\subsection{Experiment on real datasets with moving objects}

The sequence `moving car' \cite{anantrasirichai2018atmospheric} was tested. We adapted the architecture to accept 3 adjacent distorted images and 1 ground truth. Then 20 of 100 frames were used for testing whilst the rest were employed for training. In this video, the car size changes as it is coming toward us. In order to deal with different sizes of the moving car in each frame, we used random resize parameters (0.7 to 1) for data augmentation in the training process.

\begin{figure}[t!]
	\centering
  		\includegraphics[width=\columnwidth]{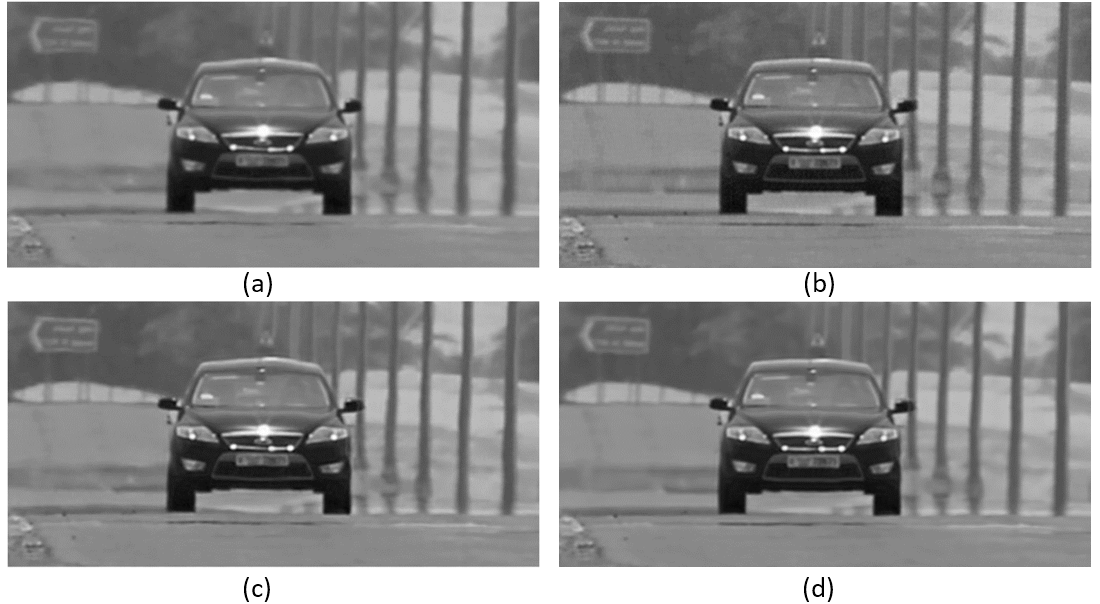}
	\caption{Reconstruction of `moving car'. (a)	distorted image, (b)`ground truth'\cite{anantrasirichai2018atmospheric} (c) directly recovered images, (d) recovered images with averaged frames.}
	\label{fig:car2}
\end{figure}

Fig. \ref{fig:car2} shows the results, where (a) is one of the distorted frame, (b) is clear frame, and (c) is its directly reconstructed image. Then, we used the average of 5  adjacent frames for each input channel (i.e. at time $t$, the inputs of each channel are frame $t$-6 to $t$-2, frame $t$-5 to $t$-1, and frame $t$-4 to $t$)  when perform testing.
The result of average input is shown in Fig. \ref{fig:car2} (d), of which the subjective quality is comparable with that of the CW  \cite{anantrasirichai2018atmospheric}  (Fig. \ref{fig:car2} (b)). 

The processing time of image restoration of this dataset (size of 512$\times$256 pixels) by using GPU is 0.08 seconds/frame, which equal to 1638.4k pixels per second. It is faster than the CW reported in \cite{anantrasirichai2018atmospheric} (70.6k pixels per second). When using CPU, the computational time for dealing the same image is 6.84 seconds, which equal to 19.2k pixels per second.

\begin{table}[t!]
	\centering
	\caption{Objective performance comparison of real dataset}
	\small
	\scalebox{1}{
		\begin{tabular}{cccccc}
		\hline
			\multirow{ 2}{*}{Image} &	\multirow{ 2}{*}{Index} & \multirow{ 2}{*}{CW} &	\multirow{ 2}{*}{CSP} &		\multicolumn{2}{c}{Proposed}  \\
			          &               &            &              &  ($\bar{y}_{20}$ ) & ($\bar{x}$) \\
			\hline
			\multirow{ 2}{*}{Chimney}	 & PSNR & 	32.02	& 31.05	 & 30.67 & 	\textbf{33.68} \\
	                            & SSIM &	 0.96	& 0.92	& 0.95	& \textbf{0.97} \\
            \hline 
             \multirow{ 2}{*}{Building}	& PSNR & 	25.18	&  25.37	&  23.23	&  \textbf{26.02} \\
	& SSIM	 & \textbf{0.91} &  	0.79	& 0.82	& 0.88 \\
				\hline 
		\end{tabular}}
				\begin{flushleft} 
				$\bar{y}_{20}$=average PSNR or SSIM of 20 recovered images \\
				$\bar{x}$=input is the average of 30 images for `chimney' and 20 for `building'
				\end{flushleft}
	\label{tab:performance}
\end{table}

\section{Conclusions}
\label{sec:conclude}

This paper presents a deep learning method with residual learning for removing atmospheric turbulence distortion using modified DnCNN. It was tested with simulated datasets, real datasets both static and dynamic scenes. We showed that the proposed model can handle turbulence with unknown distortion levels and restore image with only one distorted image, and better result can be achieved with an average of multiple distorted frames.
Our method gives equal or better results than the state-of-the-art methods without prior selecting good frames. With the simple averaging strategy, we can deal with distortions from the real work including the videos with moving objects. Moreover, it can achieve on-line mitigation for videos with static scene with GPU computation.

\newpage
\balance
\bibliographystyle{IEEEbib}

\bibliography{heathazeBib}

\end{document}